\documentclass{svproc}

\usepackage{graphicx}
\usepackage{wrapfig}
\usepackage{url}

\newcommand{\AVG}[1]{\ensuremath{\left\langle #1 \right\rangle}}

\begin{document}
\mainmatter

\title{Overview of experimental critical point search}
\titlerunning{Overview of experimental CP search}

\author{Tobiasz Czopowicz\inst{1,2}}
\authorrunning{Tobiasz Czopowicz}
\tocauthor{Tobiasz Czopowicz}

\institute{
  Jan Kochanowski University, Kielce, Poland\\
  \and
  Warsaw University of Technology, Warsaw, Poland\\
  \email{Tobiasz.Czopowicz@cern.ch}
}

\maketitle

\begin{abstract}
The existence and location of the QCD critical point is an object of vivid experimental and
theoretical studies. Rich and beautiful data recorded by experiments at SPS and RHIC allow for a
systematic search for the critical point -- the search for a non-monotonic dependence of various
correlation and fluctuation observables on collision energy and size of colliding nuclei.
\end{abstract}

\section{Critical point search strategies} 

\begin{wrapfigure}{r}{.5\textwidth}
  \centering
  \vspace{-\intextsep}
  \includegraphics[width=.45\textwidth]{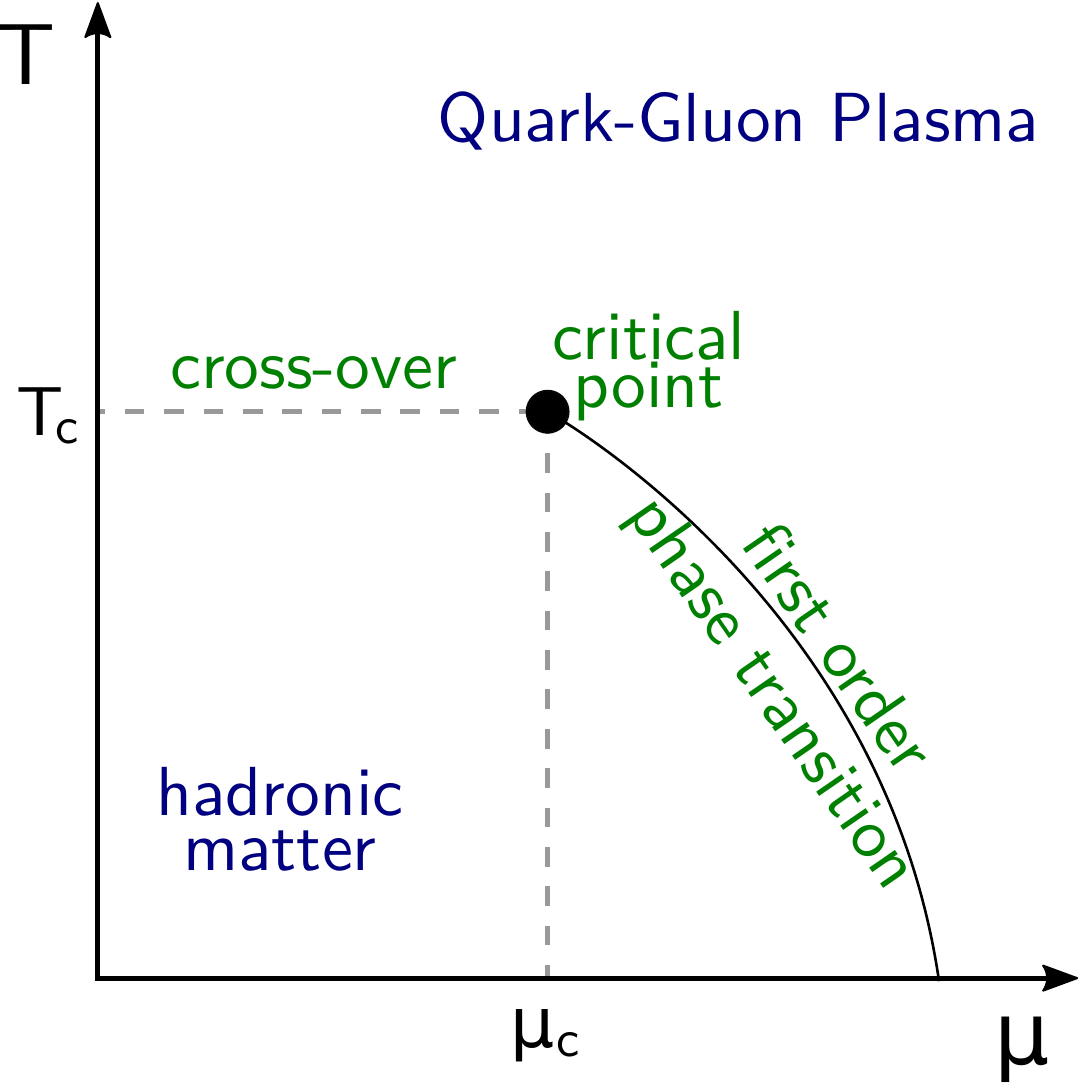}
  \caption{
    A sketch of the phase diagram of strongly-interacting matter.
  }
  \label{fig:phase-diagram}
\end{wrapfigure}

A sketch of the most popular phase diagram of strongly-interacting matter is shown in
Fig.~\ref{fig:phase-diagram}. At low temperatures and baryon chemical potential, the system consists
of quarks and gluons confined inside hadrons. At higher temperature and/or baryon chemical potential,
quarks and gluons may act like quasi-free particles, forming a different state of matter -- the
Quark-Gluon Plasma. Between the two phases, a first-order transition is expected at high $\mu$.
Critical point (CP) is a hypothetical end point of this first-order phase transition line that has
properties of a second-order phase transition~\cite{Asakawa:1989bq,Barducci:1989wi}.

It is commonly expected that the QCD critical point should lead to an anomaly in fluctuations in a
narrow domain of the phase diagram. However predictions on the CP existence, its location and what
and how should fluctuate are model dependent~\cite{Stephanov:2007fk}.

The experimental search for the critical point requires a two-dimensional scan in freeze-out
parameters (T, $\mu$) by changing collision parameters controlled in laboratory, i.e. energy and
size of the colliding nuclei (or collision centrality).

\section{Experimental measures}

\subsection{Extensive quantities}

An extensive quantity is a quantity that is proportional to the number of Wounded Nucleons (W) in
the Wounded~Nucleon~Model~\cite{Bialas:1976ed} (WNM) or to the volume (V) in the
Ideal~Boltzmann~Grand~Canonical~Ensemble (\mbox{IB-GCE}). The most popular are particle number
(multiplicity) distribution $P(N)$ cumulants:\\
$\kappa_{1} = \AVG{N}$,\\
$\kappa_{2} = \AVG{(\delta N)^{2}} = \sigma^{2}$,\\
$\kappa_{3} = \AVG{(\delta N)^{3}} = S\sigma^{3}$,\\
$\kappa_{4} = \AVG{(\delta N)^{4}} - 3\AVG{(\delta N)^{2}}^{2} = \kappa\sigma^{4}$.

\subsection{Intensive quantities}

Ratio of any two extensive quantities is independent of W~(WNM) or V~(\mbox{IB-GCE}) for an event sample
with fixed W (or V) -- it is an intensive quantity. For example:
$$
  \AVG{A} / \AVG{B} = W\cdot\AVG{a} / W\cdot\AVG{b} = \AVG{a} / \AVG{b},
$$
where $A$ and $B$ are any extensive event quantities, i.e. $\AVG{A} \sim W$, $\AVG{B} \sim W$
and $\AVG{a} = \AVG{A}$ and $\AVG{b} = \AVG{B}$ for $W = 1$. Popular examples are:\\
$\kappa_{2}/\kappa_{1} = \omega[N] = \frac{\sigma^{2}[N]}{\AVG{N}} = \frac{W \cdot \sigma^{2}[n]}{W\cdot\AVG{n}} = \omega[n]$ (scaled variance),\\
$\kappa_{3}/\kappa_{2} = S\sigma$,\\
$\kappa_{4}/\kappa_{2} = \kappa\sigma^{2}$.

\subsection{Strongly intensive quantities}

For an event sample with varying W (or V), cumulants are not extensive quantities any more. For example:
$$
  \kappa_{2} = \sigma^{2}[N] = \sigma^{2}[n]\AVG{W} + \AVG{n}^{2}\sigma^{2}[W].
$$
However, having two extensive event quantities, one can construct quantities that are independent of
the fluctuations of W (or V). Popular examples include~\mbox{\cite{Gorenstein:2011vq,Gazdzicki:2013ana}}:\\
$\AVG{K}/\AVG{\pi}$,\\
$\Delta[N,P_{T}] = (\omega[N]\AVG{P_{T}} - \omega[P_{T}]\AVG{N})/c$,\\
$\Sigma[N,P_{T}] = (\omega[N]\AVG{P_{T}} + \omega[B]\AVG{N} - 2(\AVG{NP_{T}} - \AVG{P_{T}}\AVG{N})/c$,\\
where $P_{T} = \sum\limits_{i=1}^{N} p_{T,i}$ and $C$ is any extensive quantity (e.g. $\AVG{N}$).

\subsection{Short-range correlations}

Quantum statistics leads to short-range correlations in momentum space, which are sensitive to
particle correlations in configuration space (e.g. of CP origin).

Popular measures include momentum difference in Longitudinal Comoving System (LCMS), $\mathbf{q}$,
that is decomposed into three components:
$q_{long}$ -- denoting momentum difference along the beam,
$q_{out}$ -- parallel to the pair transverse-momentum vector ($\mathbf{k}_{t} =
(\mathbf{p}_{T,1} + \mathbf{p}_{T,2})/2$) and
$q_{side}$ -- perpendicular to $q_{out}$ and $q_{long}$.
The two-particle correlation function $C(q)$ is often approximated by a three-dimensional Gauss function:
$$
  C(\mathbf{q}) \cong 1 + \lambda \cdot \exp\left( -R^{2}_{long}q^{2}_{long} - R^{2}_{out}q^{2}_{out}
  - R^{2}_{side}q^{2}_{side}\right),
$$
where $\lambda$ describes the correlation strength and $R_{out}, R_{side}, R_{long}$ denote Gaussian
HBT radii.

A more parametrization of the correlation function is possible via introducing L\'{e}vy-shaped
source (1-D)~\cite{Csorgo:2005it}:
$$
  C(q) \cong 1 + \lambda \cdot e^{(-qR)^{\alpha}},
$$
where $q = |p_{1}-p_{2}|_{LCMS}$, $\lambda$ describes correlation length, $R$ determines the length
of homogenity and L\'evy exponent $\alpha$ determines source shape:\\
$\alpha = 2$: Gaussian, predicted from a simple hydro,
$\alpha < 2$: anomalous diffusion, generalized central limit theorem,
$\alpha = 0.5$: conjectured value at the critical point.

\subsection{Fluctuations as a function of momentum bin size}

When a system crosses the second-order phase transition, it becomes scale invariant, which leads to
power-law form of correlation function. The second factorial moment is calculated as a function of
the momentum cell size (or bin number $M$):
$$
  F_{2}(M) \equiv
  \bigg\langle \frac{1}{M} \sum\limits_{i=1}^{M}n_{i}(n_{i}-1)\bigg\rangle\Bigg/
  \bigg\langle \frac{1}{M} \sum\limits_{i=1}^{M}n_{i}\bigg\rangle,
$$
where $n_{i}$ is particle multiplicity in cell $i$.

At the second-order phase transition the system is a simple fractal and the factorial moment
exhibits a power-law dependence on $M$~\cite{Wosiek:APPB,Bialas:1990xd,Bialas:1985jb,Antoniou:2006zb}:
$$
F_{2}(M) \sim (M)^{\varphi_{2}}.
$$
In case the system freezes-out in the vicinity of the critical point, $\varphi_{2} = 5/6$.

To cancel the $F_{2}(M)$ dependence on the single-particle inclusive momentum distribution,
one needs a uniform distribution of particles in bins or subtraction of the $F_{2}(M)$ values for
mixed events:
$$
  \Delta F_{2}(M) =  F_{2}^{data}(M) - F_{2}^{mixed}(M).
$$

\subsection{Light nuclei production}

Based on coalescence model, particle ratios of light nuclei are sensitive to the nucleon
density fluctuations at kinetic freeze-out and thus to CP. In the vicinity of the critical point or
the first-order phase transition, density fluctuation becomes larger
\cite{Sun:2017xrx,Shuryak:2019acz}.

Nucleon density fluctuation can be expressed by proton, triton and deuteron yields as:
$$
  \Delta n = \frac{\AVG{(\delta n)^{2}}}{\AVG{n}} \approx \frac{1}{2\sqrt{3}}\frac{N_{p}\cdot
  N_{t}}{N_{d}^{2}}-1.
$$

\section{Experimental results} 

\subsection{Multiplicity fluctuations}

\begin{figure}[!htb]
  \centering
  \vspace{0pt}\includegraphics[width=.33\textwidth]{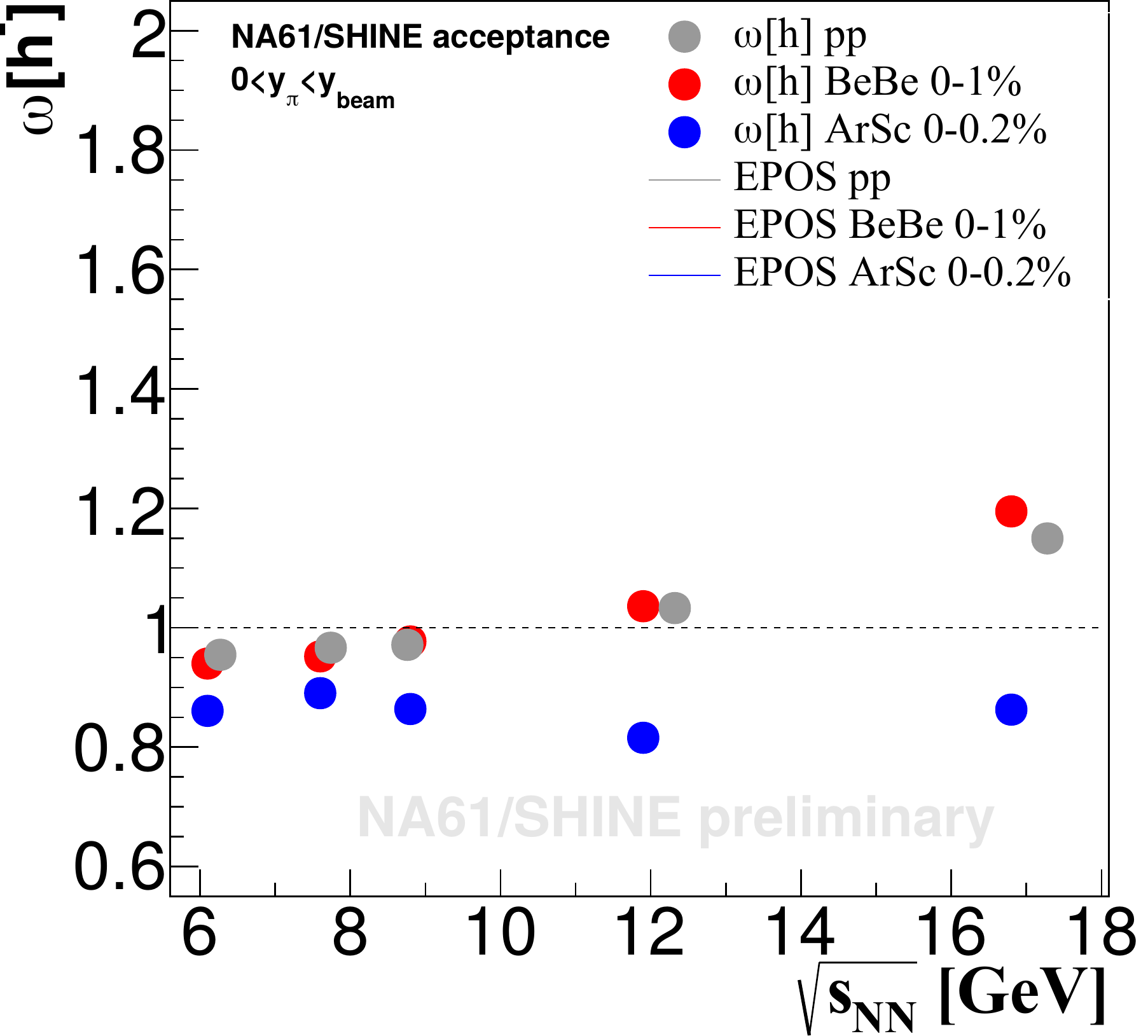}\hfill
  \vspace{0pt}\includegraphics[width=.32\textwidth]{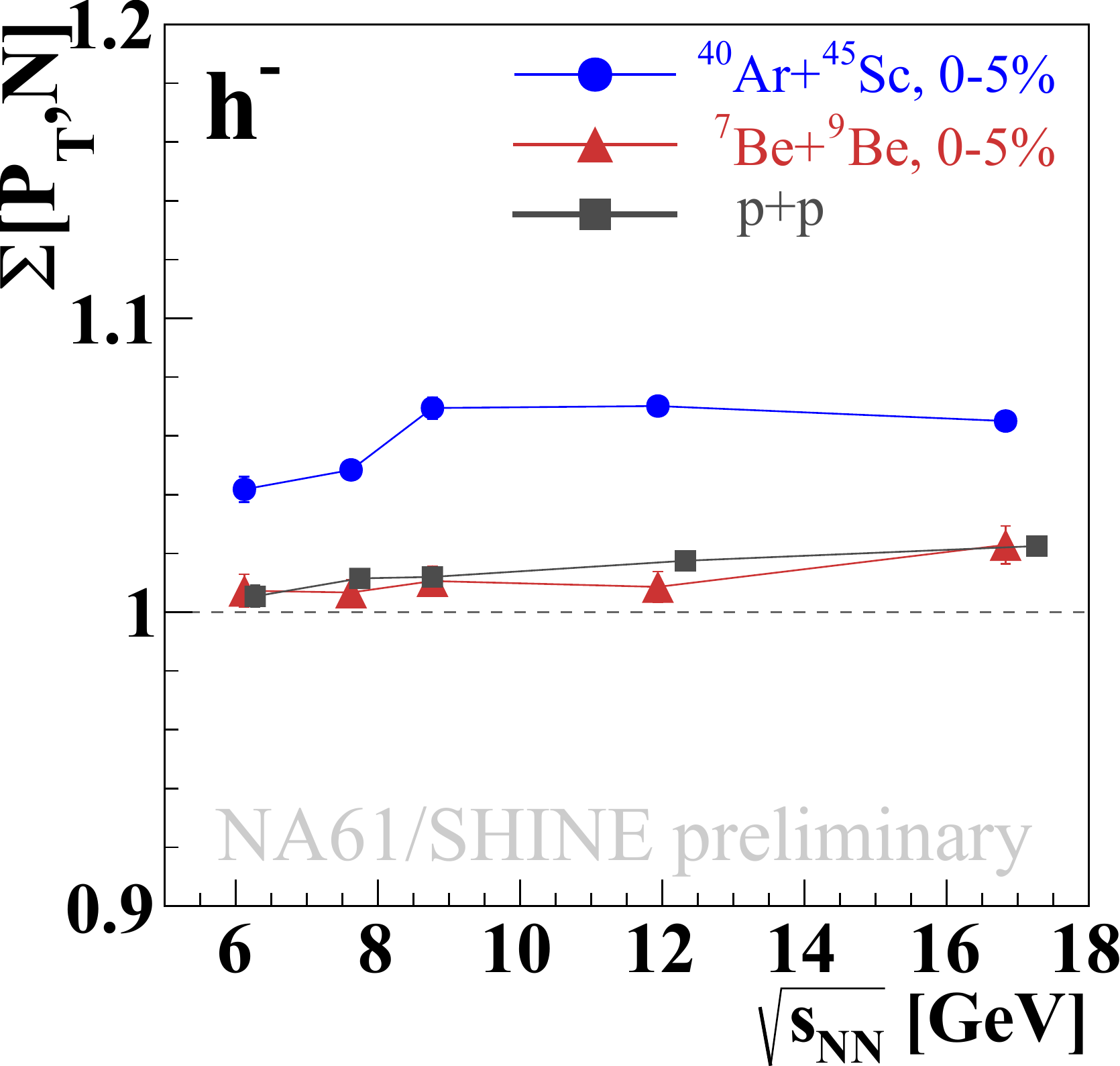}\hfill
  \vspace{0pt}\includegraphics[width=.32\textwidth]{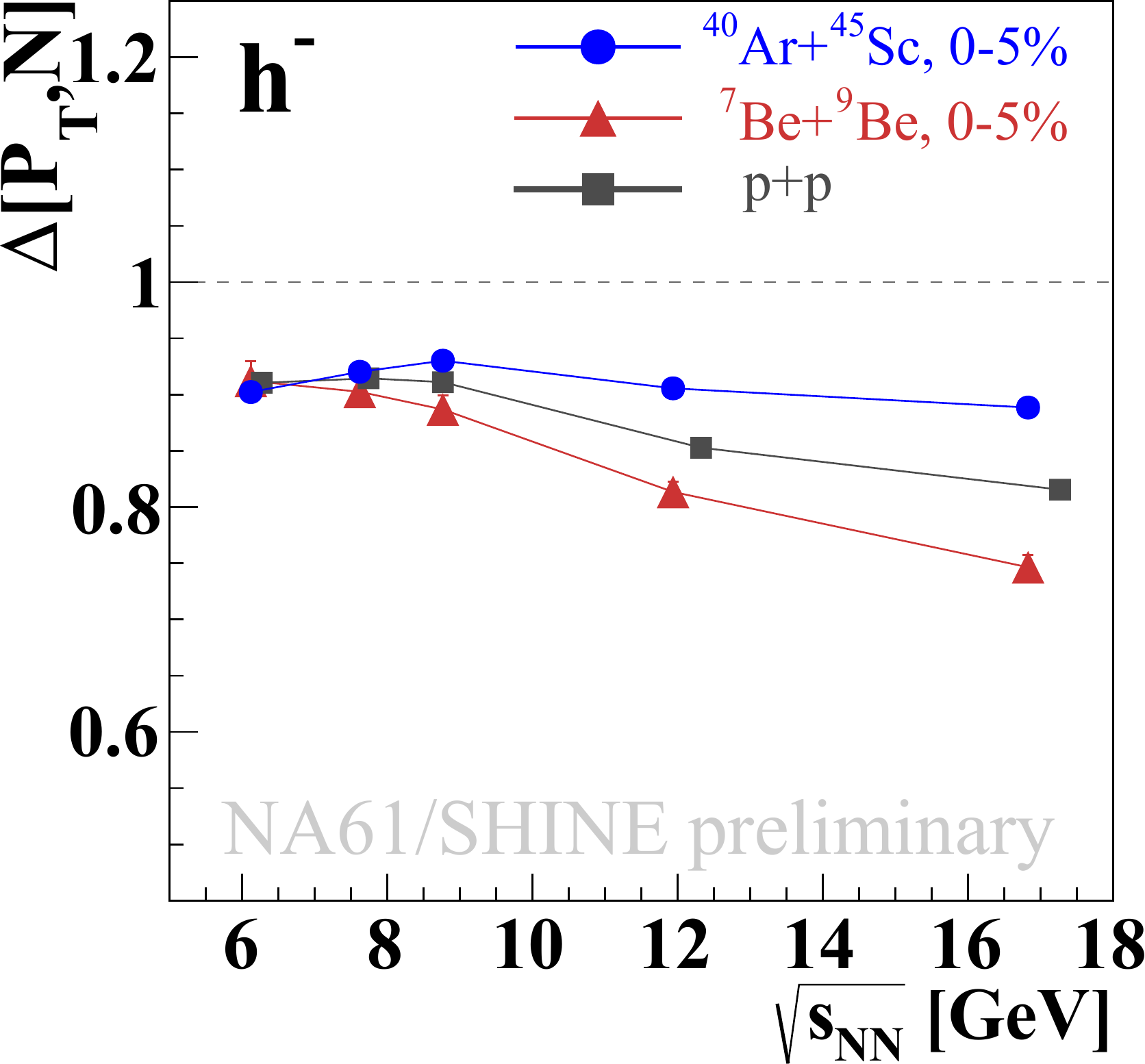}
  \caption{
    Results on multiplicity~\cite{Gazdzicki:2017zrq} ({\it left}) and multiplicity-transverse
    momentum~\cite{Andronov:2016ddd} ({\it center, right}) fluctuations for all negatively charged
    particles recorded by NA61/SHINE.
  }
  \label{fig:mult-pt-fluct}
\end{figure}

Results on energy dependence of multiplicity fluctuations by NA61/SHINE~\cite{Gazdzicki:2017zrq}
quantified by the scaled variance are presented in Fig.~\ref{fig:mult-pt-fluct} ({\it left}).
No prominent structures that could be related to the critical point are observed.

\subsection{Multiplicity-transverse momentum fluctuations}

Results on energy dependence of multiplicity-transverse momentum fluctuations by NA61/SHINE
\cite{Andronov:2016ddd} expressed in $\Delta$ and $\Sigma$ strongly intensive quantities are
presented in Fig.~\ref{fig:mult-pt-fluct} ({\it center, right}).
No prominent structures that could be attributed to the critical point are observed.

\subsection{Net-proton fluctuations}

\begin{figure}[!htb]
  \centering
  \includegraphics[width=.48\textwidth]{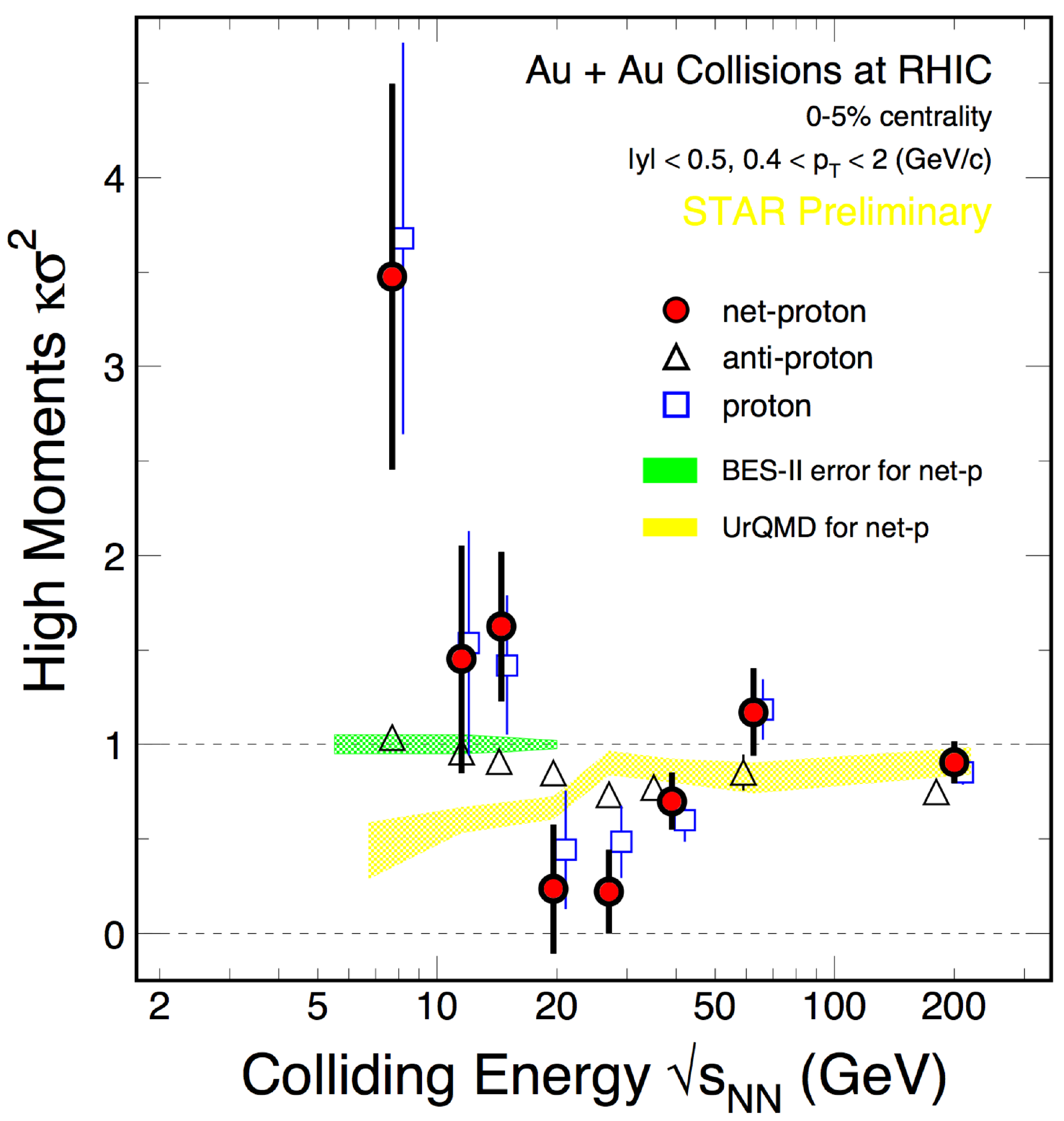}\hfill
  \includegraphics[width=.46\textwidth]{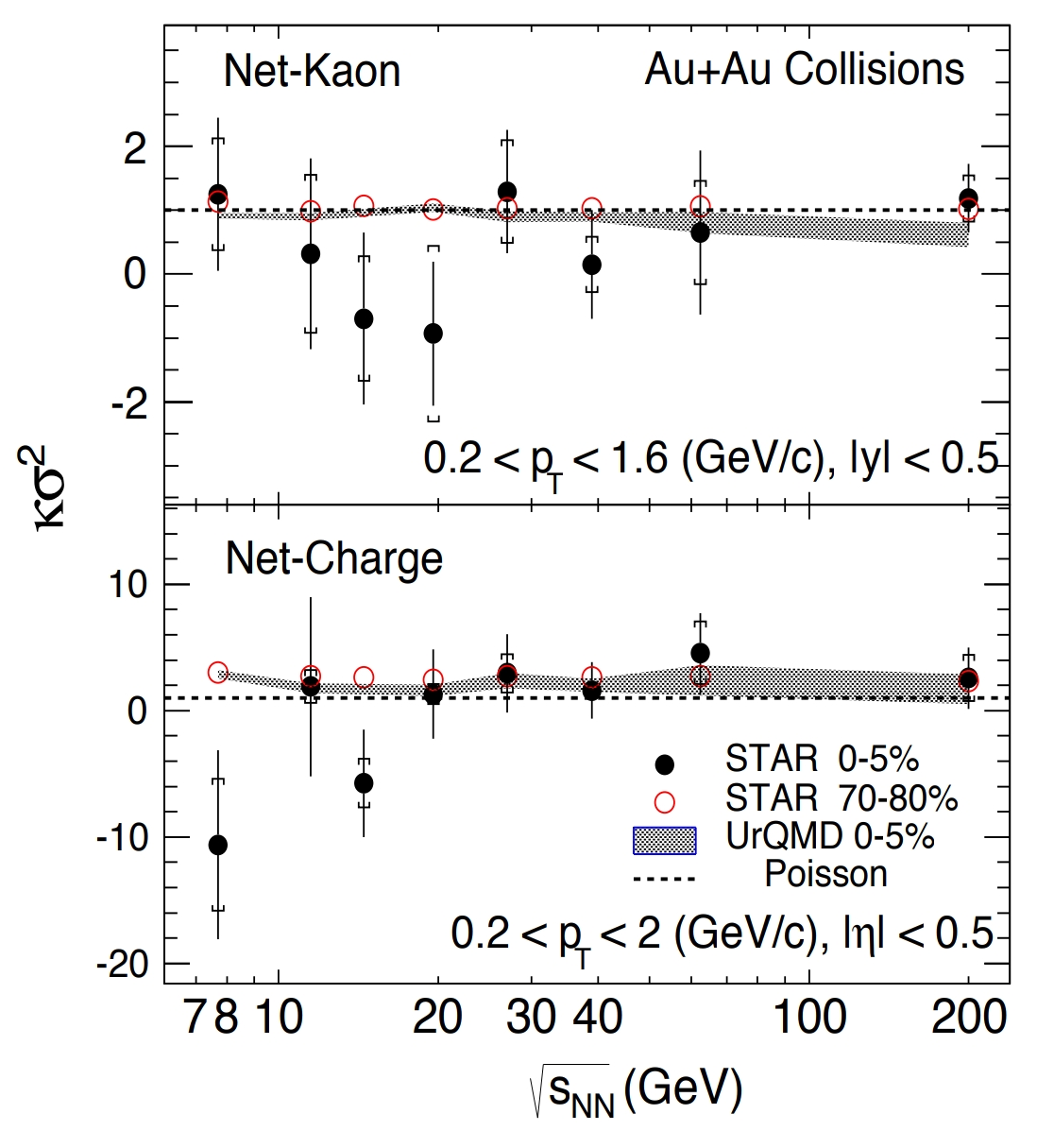}
  \caption{
    Results on $\kappa\sigma^{2}$ of net-proton~\cite{Adamczyk:2013dal} ({\it left}) as well as net-kaon
    and net-charge~\mbox{\cite{Adamczyk:2017wsl,Adamczyk:2014fia}} ({\it right}) distributions measured by
    STAR.
  }
  \label{fig:net-p-k-charge}
\end{figure}

Figure~\ref{fig:net-p-k-charge} ({\it left}) presents energy dependence of fourth-order net-proton
fluctuation in 5\% most central Au+Au collisions recorded by STAR~\cite{Adamczyk:2013dal}.
The observed non-monotonic dependence is consistent with theoretical predictions~\cite{Stephanov:2011pb}
and might suggest a critical point around $\sqrt{s_{NN}} \approx 7$ GeV.

\subsection{Net-kaon and net-charge fluctuations}

The STAR Collaboration has also studied net-kaon and net-charge distributions in central Au+Au
collisions~\cite{Adamczyk:2017wsl,Adamczyk:2014fia}.
However, the results, presented in Fig.~\ref{fig:net-p-k-charge} ({\it right}), show
no (within errors) energy dependence.

\subsection{Short-range correlations}

\subsubsection{Finite-Size Scaling}

\begin{figure}[!htb]
  \centering
  \includegraphics[width=.40\textwidth]{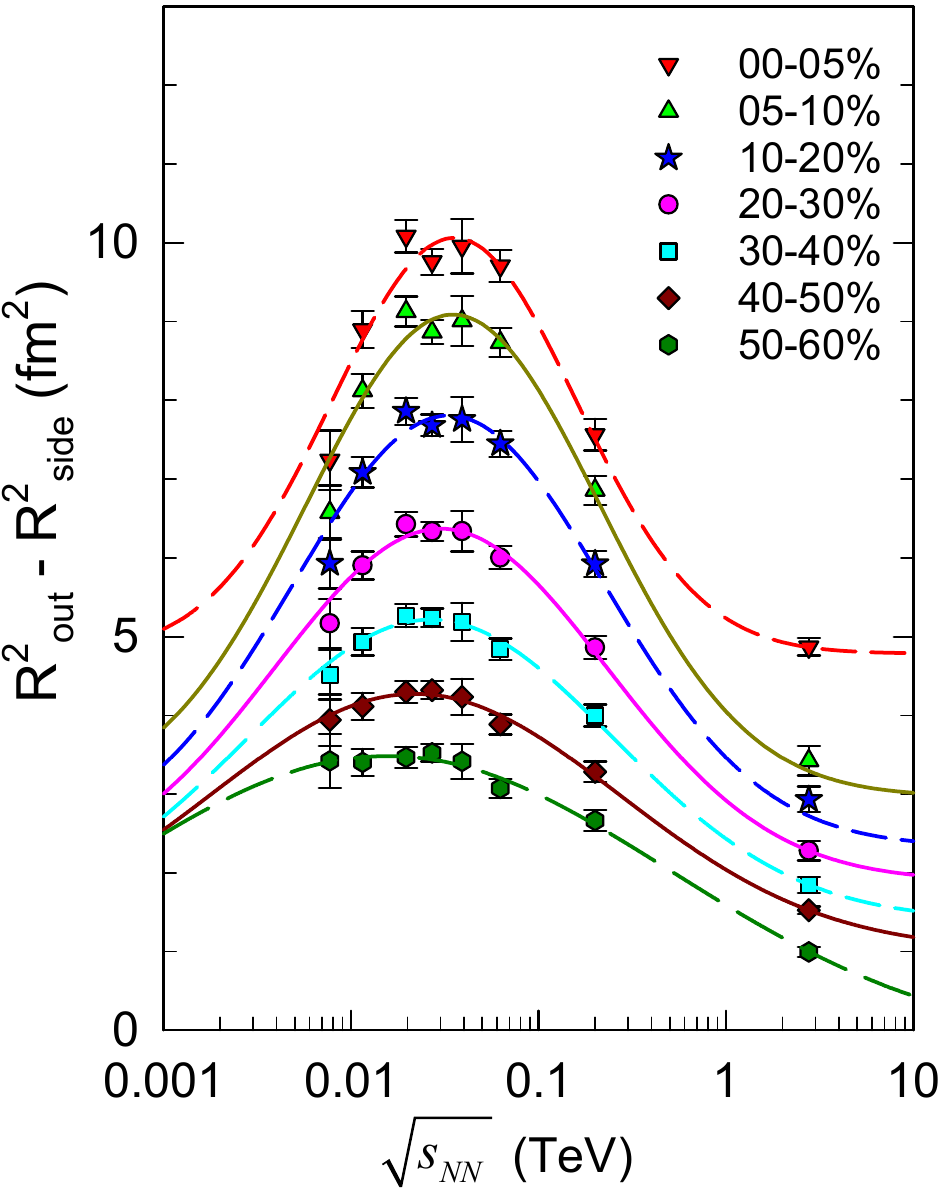}\hfill
  \includegraphics[width=.52\textwidth]{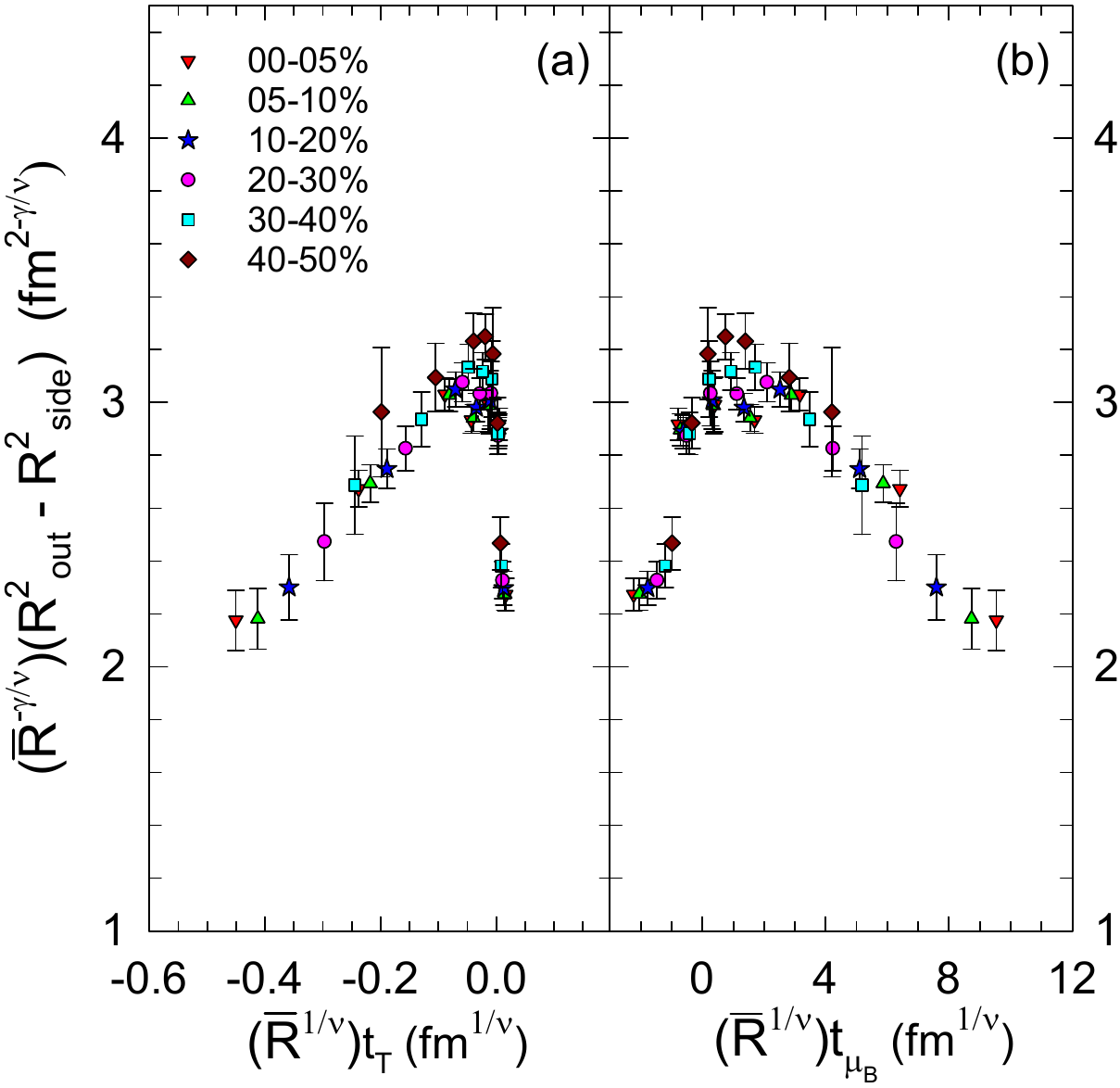}
  \caption{
    Compilations of Au+Au ($\sqrt{s_{NN}}$ = 7.7--200 GeV, STAR~\cite{Adamczyk:2014mxp}) and Pb+Pb
    ($\sqrt{s_{NN}} = 2.76$ TeV, ALICE~\cite{Aamodt:2011mr}) data: energy dependence of Gaussian
    emission source radii~\cite{Lacey:2014wqa} ({\it left}) and one of the result for initial
    Finite-Size Scaling analysis~\cite{Lacey:2014wqa} ({\it right}).
  }
  \label{fig:lacey}
\end{figure}

Fig.~\ref{fig:lacey} presents compilations of Au+Au (\mbox{$\sqrt{s_{NN}}$ = 7.7--200} GeV) data from
STAR~\cite{Adamczyk:2014mxp} and Pb+Pb ($\sqrt{s_{NN}} = 2.76$ TeV) data from ALICE
\cite{Aamodt:2011mr}.
The Gaussian emission source radii ($R^{2}_{out} - R^{2}_{side}$)~\cite{Lacey:2014wqa} show clear
non-monotonic energy dependence  with a maximum at $\sqrt{s_{NN}} \approx 47.5$ GeV.
The initial Finite-Size Scaling analysis~\cite{Lacey:2014wqa} suggests the critical point position:
\mbox{$T = 165$ MeV} and $\mu = 95$ MeV.

\subsubsection{Transverse-mass dependence of L\'{e}vy exponent}

\begin{figure}[!htb]
  \centering
  \includegraphics[width=0.49\textwidth]{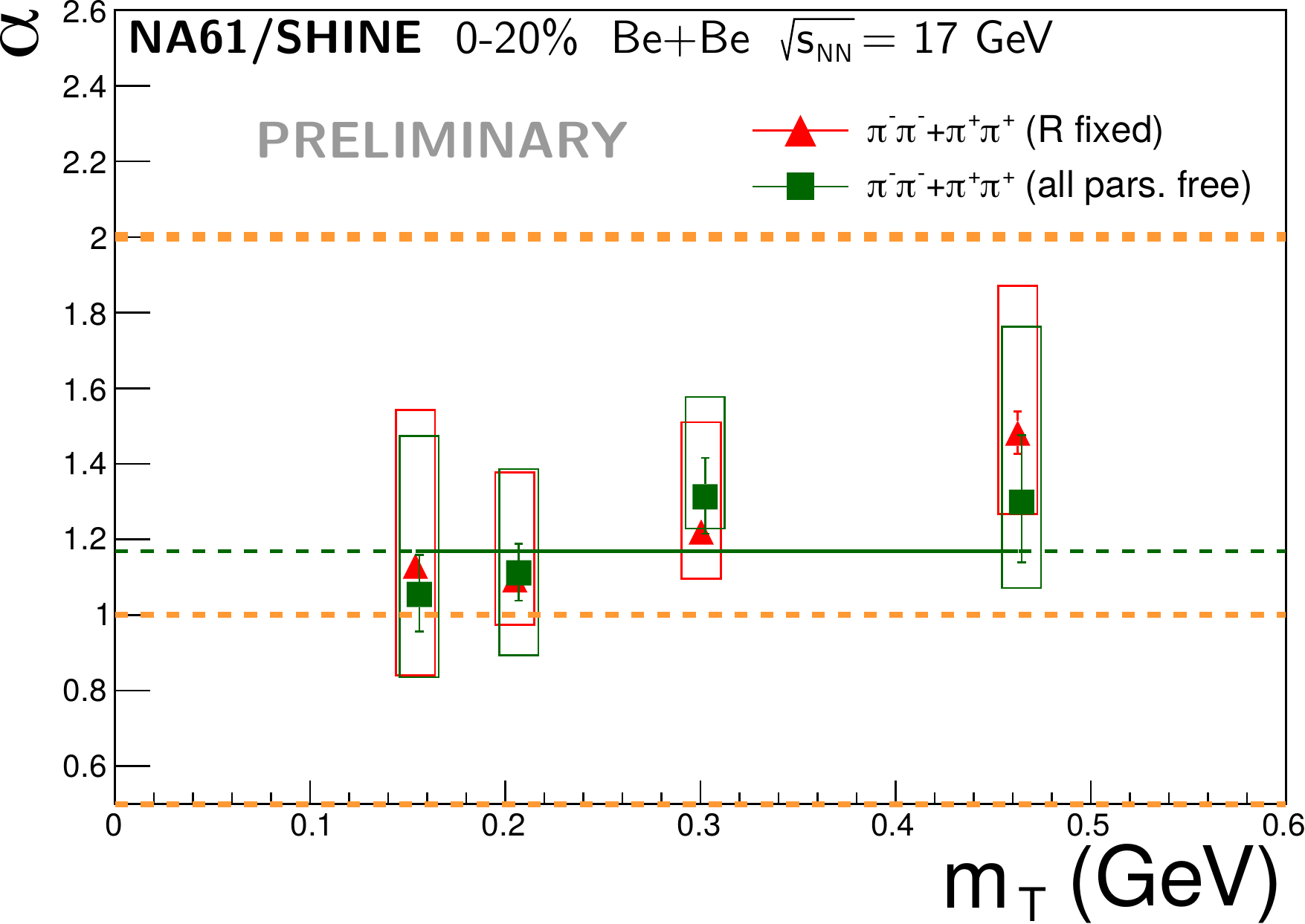}\hfill
  \includegraphics[width=0.47\textwidth]{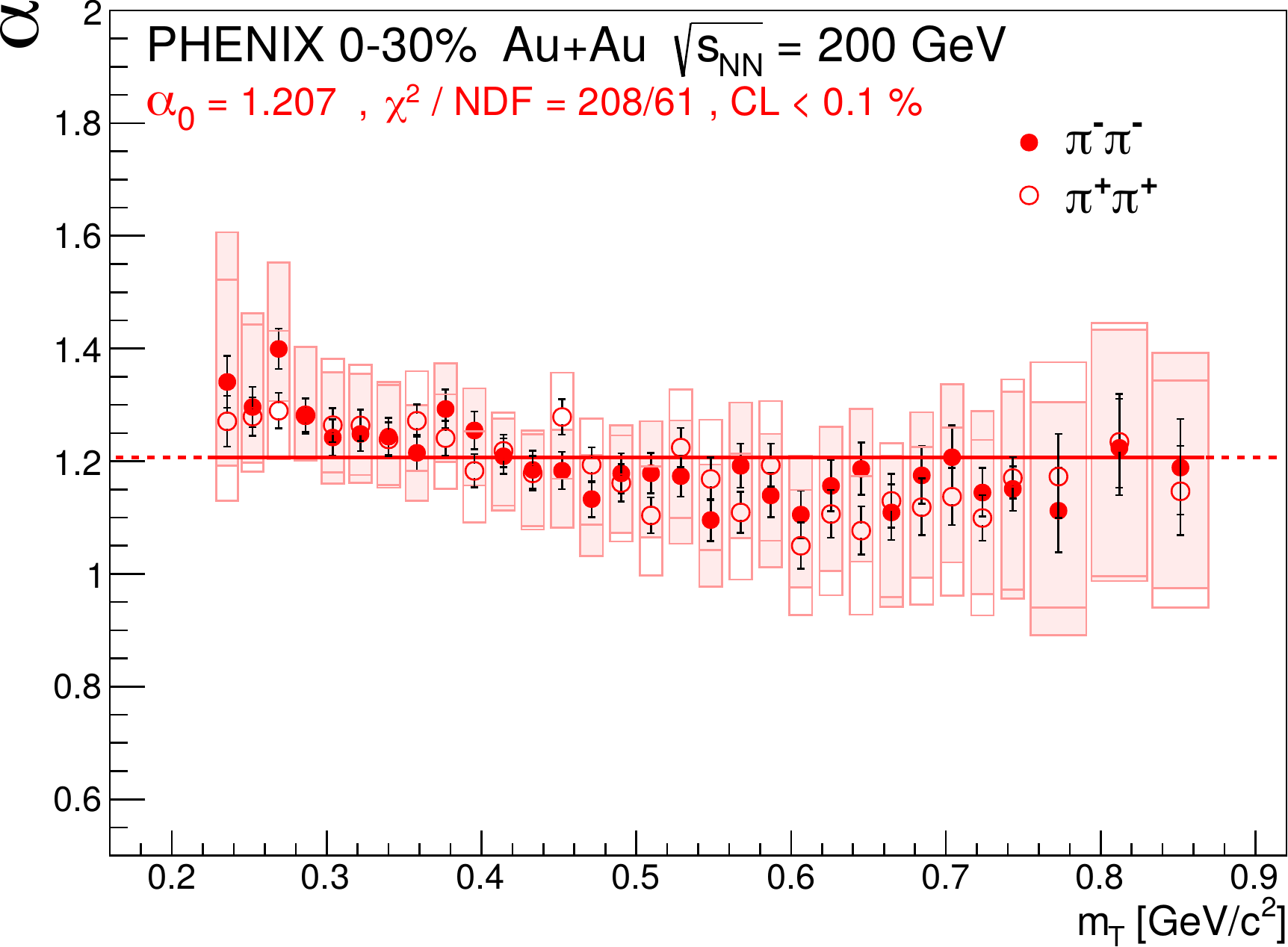}
  \caption{
    Transverse mass dependence of the L\'{e}vy exponent $\alpha$ for 20\% most central Be+Be collisions
    at 17 GeV by NA61/SHINE~\cite{Porfy:2018mvd} ({\it left}) and for 30\% Au+Au at 200 GeV by
    PHENIX~\cite{Adare:2017vig} ({\it right}).
  }
  \label{fig:levy}
\end{figure}

Transverse-mass dependence of L\'{e}vy exponent $\alpha$ have been studied both at SPS and RHIC.
Figure~\ref{fig:levy} presents the results for Be+Be at 17 GeV by NA61/SHINE~\cite{Porfy:2018mvd}
and for Au+Au at 200 GeV by PHENIX~\cite{Adare:2017vig}.
Both studies revealed similar results, i.e. $\alpha \approx 1.2$, a value significantly above
the CP prediction.

\subsection{Fluctuations as a function of momentum bin size}

NA49 and NA61/SHINE have studied the second factorial moment, $\Delta F_{2}$, for
mid-rapidity protons at 17 GeV.

\begin{figure}[!htb]
  \centering
  \includegraphics[width=.32\textwidth]{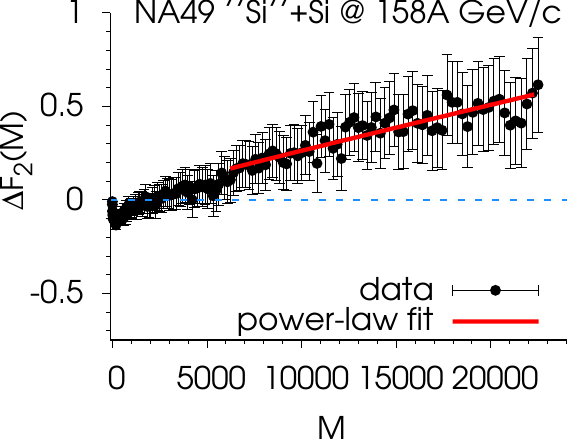}\hfill
  \includegraphics[width=.32\textwidth]{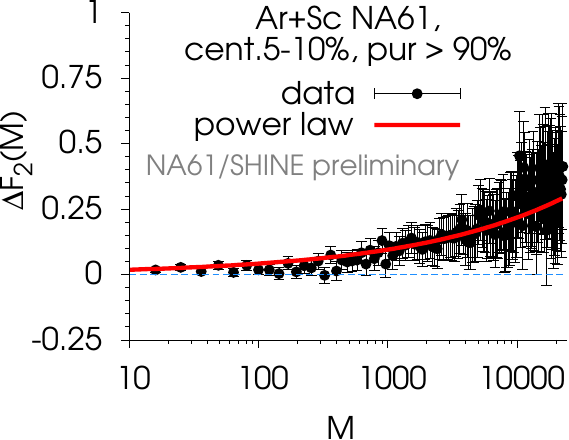}\hfill
  \includegraphics[width=.32\textwidth]{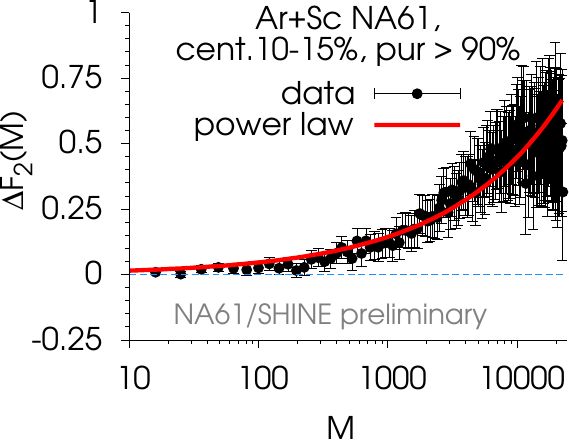}
  \caption{
    Second factorial moment, $\Delta F_{2}$, for mid-rapidity protons at 17 GeV in Si+Si by NA49
   ~\cite{Anticic:2012xb} ({\it left}) and in 5--10\% and 10--15\% Ar+Sc by NA61/SINE
   ~\cite{Davis:2017puj} ({\it center, right}).
  }
  \label{fig:davis}
\end{figure}

Although in central Be+Be, C+C, Ar+Sc and Pb+Pb no signal has been observed,
a deviation of $\Delta F_{2}$ from zero seems apparent in central Si+Si and mid-central Ar+Sc as
shown in Fig.~\ref{fig:davis}.

\subsection{Light nuclei production}

\begin{figure}[!htb]
  \centering
  \includegraphics[width=0.5\textwidth]{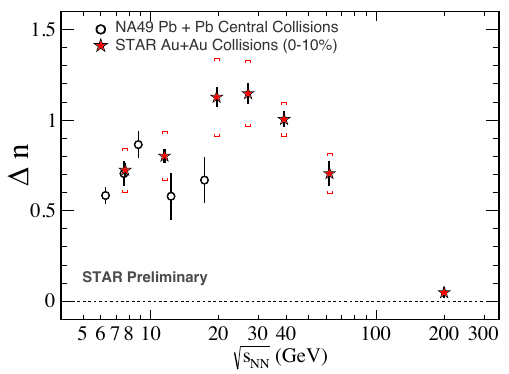}
  \caption{
    Nucleon density fluctuation, $\Delta n$, for central Pb+Pb~\cite{Anticic:2016ckv} and
    Au+Au~\cite{Adamczyk:2017nof,Yu:2017bxv} collisions.
  }
  \label{fig:light-nuclei}
\end{figure}

The nucleon density fluctuations, $\Delta n$, for central Pb+Pb by NA49~\cite{Anticic:2016ckv} and
central Au+Au by STAR~\cite{Adamczyk:2017nof,Yu:2017bxv} show a non-monotonic dependence on collision
energy with a peak for $\sqrt{s_{NN}} \approx 20$ GeV~\cite{Zhang:2017bxv} as presented in
Fig.~\ref{fig:light-nuclei}.

\section{Summary}

\begin{figure}[!htb]
  \centering
  \includegraphics[width=0.7\textwidth]{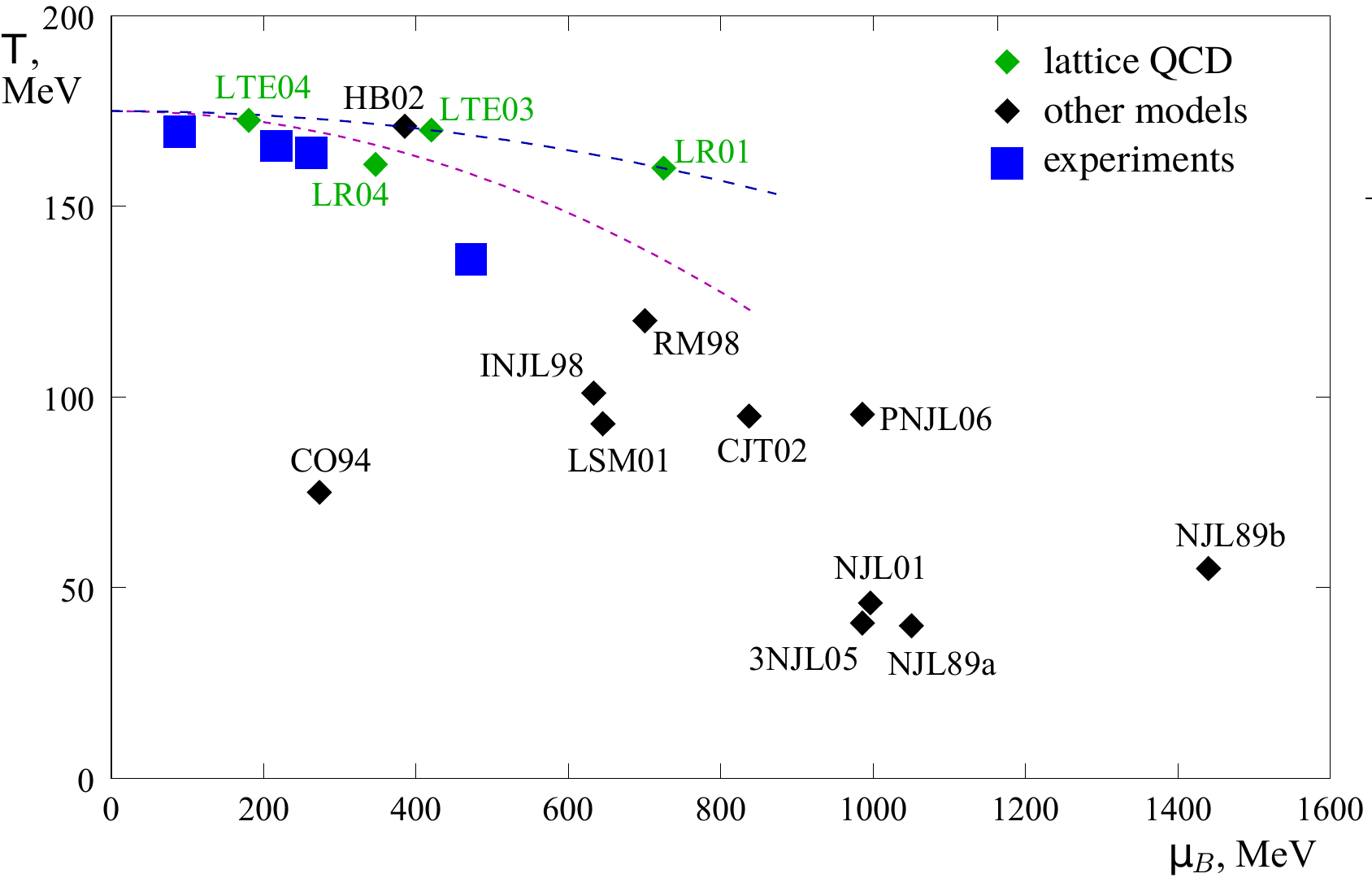}
  \caption{
    Compilation of theoretical predictions~\cite{Stephanov:2007fk} and experimental hints on the
    critical point location.
  }
  \label{fig:summary}
\end{figure}

The experimental search for the critical point is ongoing. There are four indications of anomalies
in fluctuations in heavy-ion collisions at different collision energies ($\sqrt{s_{NN}} \approx 7, 17, 20, 47$ GeV).
Interpreting them as due to CP allows one to estimate four hypothetical CP locations depicted in
Fig.~\ref{fig:summary}.

Fortunately, there are high-quality, beautiful new data coming soon both from SPS
(\mbox{NA61/SHINE}) and RHIC (STAR Beam energy Scan II).

\end{document}